\documentclass[letterpaper, 10 pt, conference]{ieeeconf}
\IEEEoverridecommandlockouts                              
\makeatletter
\let\NAT@parse\undefined
\makeatother

\usepackage[dvipsnames]{xcolor}

\newcommand*\linkcolours{ForestGreen}

\usepackage{times}
\usepackage{graphicx}
\graphicspath{{./figs/}}
\usepackage{amssymb}
\usepackage{gensymb}
\usepackage{amsmath}
\usepackage{hyperref}
\usepackage{breakurl}

\usepackage{url,hyperref}
\hypersetup{
colorlinks,
linkcolor=\linkcolours,
citecolor=\linkcolours,
filecolor=\linkcolours,
urlcolor=\linkcolours}

\usepackage{algorithm}
\usepackage{algorithmic}

\usepackage[labelfont={bf},font=small]{caption}
\usepackage[none]{hyphenat}

\usepackage{mathtools, cuted}

\usepackage[noadjust, nobreak]{cite}

\usepackage{tabularx}
\usepackage{amsmath}

\usepackage{float}

\usepackage{pifont}

\newcolumntype{Y}{>{\centering\arraybackslash}X}

\usepackage[]{placeins}

\usepackage{placeins}

\usepackage{tikz}

\usepackage[framemethod=tikz]{mdframed}

\usepackage{afterpage}

\usepackage{stfloats}

\usepackage{atbegshi}
\newcommand{\handlethispage}{}
\newcommand{\discardpagesfromhere}{\let\handlethispage\AtBeginShipoutDiscard}
\newcommand{\keeppagesfromhere}{\let\handlethispage\relax}
\AtBeginShipout{\handlethispage}

\usepackage{comment}
\usepackage{authblk}

\title{\LARGE \bf
Evaluation and Comparison of Edge-Preserving Filters
}

\author{Sarah Gingichashvili and Dani Lischinski\\\\School of Computer Science and Engineering\\The Hebrew University of Jerusalem, Israel}    

\begin{document}
\maketitle
\thispagestyle{empty}
\pagestyle{empty}

\begin{abstract}
Edge-preserving filters play an essential role in some of the most basic tasks of computational photography, such as abstraction, tonemapping, detail enhancement and texture removal, to name a few. The abundance and diversity of smoothing operators, accompanied by a lack of methodology to evaluate output quality and/or perform an unbiased comparison between them, could lead to misunderstanding and potential misuse of such methods. This paper introduces a systematic methodology for evaluating and comparing such operators and demonstrates it on a diverse set of published edge-preserving filters. Additionally, we present a common baseline along which a comparison of different operators can be achieved and use it to determine equivalent parameter mappings between methods. Finally, we suggest some guidelines for objective comparison and evaluation of edge-preserving filters.
\end{abstract}

\section{Introduction}

Edge-aware filters are among the most commonly used image processing operators in a wide range of computational photography applications. Also called \emph{edge-preserving smoothing} operators, they form a pivotal step in many algorithms for image manipulation including tonemapping, image abstraction, detail enhancement, texture removal, and more. The ubiquity of these operators may be attributed to their seeming ability to identify salient, structure-defining edges in an image and to process them differently from other image details and discontinuities. 

Over the past decade, a large number of edge-preserving smoothing operators were introduced in the literature, some of which are explicitly designed for specific tasks, such as texture removal, or multiscale detail manipulation, while others without an explicitly declared goal. We mention several specific examples in Section \ref{sec-smoothing-operators}.

These operators differ from each other in several aspects, such as their visual style, internal algorithmics (e.g., local vs.~global operators), the number and range of parameters (from at least one to many), and runtime complexity (seconds to minutes). Such differences may not be readily apparent to a user who is only casually familiar with these methods and, therefore, the task of selecting the right operator for a specific purpose (or a specific image), as well as choosing its parameters, can be quite challenging.

When introducing a new filter, researchers typically include side-by-side visual comparisons of images processed by existing operators. In order for such a comparison to be meaningful, an image should be processed by the different operators to an \emph{equivalent degree}. However, each operator is controlled by its own parameter(s), and there is no established consistent mapping between the parameters of different operators. This is further complicated by the fact that even for each single method the visual effect of its parameters on the result is often input-dependent, and the relationship between its parameters could vary at different smoothing levels. Thus, in practice, the parameters for each of the compared methods are chosen subjectively by the authors, typically through trial-and-error.

The choice of which methods to compare to is also nontrivial. Ideally, one would like to include in the comparison methods related in their approach to the proposed one, alongside with ones representing completely different approaches. It is also important to take into account the main intended application for which the competing methods have been originally designed. Finally, a common benchmark should be used for all comparisons, but, in practice, no established benchmark exists; furthermore, it is not uncommon to see cases where a different image is chosen to compare the proposed method to each of the existing competitors.

In this paper, we propose a new methodology for evaluating and comparing edge-preserving smoothing operators in a more meaningful, objective, and quantitative fashion. We propose a mechanism for achieving roughly equivalent smoothing degrees by different operators, which effectively establishes a mapping between their main governing parameters, on a per-image basis. Having established such a mapping we can quantitatively measure various characteristics of different operators. This provides several insights on the operators' behavior, and enables performing more meaningful comparisons, both visually and quantitatively. Finally, we are able to cluster the different operators based on the similarity of their effect, using strictly empirical observations. Our observations on innate method behavior suggest some specific recommendations for both developers and end-users of such operators.

The remainder of this paper is organized as follows: Section~\ref{sec-background} reviews existing approaches for comparing between tonemapping operators and several previous comparative surveys of edge-preserving filters. Section~\ref{sec-methods} first briefly lists the filters that we chose to demonstrate our methodology for comparing and evaluating smoothing operators, and then introduces the methodology itself. Section~\ref{sec-results} demonstrates the use of our methodology on the selected set of operators. Section~\ref{sec-conclusions} provides a concluding discussion and offers some ideas for future research.
\section{Background}\label{sec-background}

Edge-preserving smoothing of images may be achieved by a wide variety of numerical approaches originating with nonlinear diffusion methods in the early 90's \cite{perona1990scale}, through bilateral filtering \cite{tomasi1998bilateral}, to modern approaches such as edge-avoiding wavelets by \cite{fattal2009edge}. Despite the relative maturity of this area, new algorithms continue to be published to this day, e.g.,  \cite{tang2017effective,ham2017robust}. Due to the subjective nature of edge-preserving filters (namely, in judging which edges “deserve” to be preserved in an image), there are no established criteria or sets of attributes according to which smoothing operators are evaluated. In contrast, a number of papers were published in recent years on the related topic of comparison and evaluation of tone mapping operators (TMOs). These works tend to focus on one of the approaches discussed below. 

\subsection{Subjective perceptual evaluations}

A popular approach for evaluating TMOs is via subjective psychophysical experiments in which human observers are typically asked to give an opinion on the quality of a given tonemapped image. In these cases, the subjects are often asked to rank images according to their preference or a predefined list of subjective features. Examples where such an approach was utilized include work by \v{C}ad\'{i}k et al. ~\cite{vcadik2006image}, who use image attributes such as perceived brightness, apparent contrast, visibility of detail, and reproduction of color. These attributes are used to define an overall image quality measure based on two experiments, in which subjects were asked to: (a) rate tonemapped images given a ground truth HDR image, and (b) rank tonemapped images with no reference to the ground truth. Another, more recent example is the work of Eilertsen et al.~\cite{eilertsen2013evaluation}, who also perform a user-study for HDR video tonemapping evaluation, in which observers provide a direct rating and pairwise comparison results of operators based on image attributes such as color saturation and flickering. 

\subsection{Partial ground truth approximation}

In another approach, the closeness to a (non-existing) ground truth is approximated by statistical metrics of natural scenes. Kundu et al.~\cite{kundu2016no} assess tonemapped image quality by performing comparisons to known natural scene statistics, such as mean subtracted contrast normalized (MSCN) pixels and log-derivative statistics. This approach is common in no-reference image quality assessment tasks, where other types of statistical metrics have also been used, e.g., image gradient statistics \cite{liang2010no} and discrete cosine transform (DCT) statistics \cite{brandao2008no}.

\subsection{Objective feature-based evaluations }

Yet another approach attempts to bridge subjective measures with an objectively measurable set of image features. Kumar et al.~\cite{kumar2017no} train a convolutional neural network (CNN) based on a 4096-length feature vector to perform quality assessment of tonemapped HDR images via a linear regression model fit to crowdsourced Mean Opinion Scores (MOS).

In contrast, significantly less work was done on the evaluation and comparison of smoothing operators. This is in large part due to a lack of consensus on what constitutes a “well-smoothed” image: the answer is effectively determined by image semantics as they are interpreted by the observer (e.g., what is the optimal tradeoff between preserved edges and smoothed areas in the image). For this reason, subjective psychophysical experiments would not be as effective at comparing smoothing operators as they are at comparing TMOs. Moreover, while one can approximate a ground truth or a “well-tonemapped” image by evaluating its “naturalness” or by simply viewing it on an HDR display, the same cannot be done for edge-aware smoothing. Indeed, since the results are not meant to look “realistic”, one cannot evaluate them as such. Finally, the result of an evaluation depends on the task at hand. For example, an operator that produces flat images with aggressive edge elimination may be preferred for abstraction purposes, but may be not be desirable for tasks such as detail enhancement. 

Nonetheless, several previous attempts were made to review edge-preserving smoothing operators. Goyal et al. \cite{goyal2012comprehensive} provide a brief overview of the main approaches to image smoothing methods. Pal et al. \cite{pal2015brief} survey several smoothing algorithms, while focusing on the interrelationships between various mathematical foundations of edge preserving filters. 
While such surveys are helpful for those who wish to get an understanding of the design and the mathematical foundations of different approaches, they provide little practical guidance in terms of the algorithms' output style and the range and effect of their input parameters on the output. In particular, there is no consensus on what constitutes equivalently smoothed images by different edge-preserving filters. Our methodology, which is similar to the aforementioned approaches to evaluating tonemapping operators in that it is image attribute-based, allows for such an equivalency to be defined.
\section{Methodology}\label{sec-methods}

In this section, we introduce our methodology by first listing the edge-preserving smoothing operators that were chosen to demonstrate it (\ref{sec-smoothing-operators}). For each method we identify its primary controlling parameter, along with corresponding empirically determined maximal values (\ref{sec-primary-parameters}). The methods are analyzed and compared to each other in Section \ref{sec-results} by measuring their effect on several pertinent image attributes, which are described in Section \ref{sec-img-attributes}. One of these attributes is gradient attenuation, which we also use in order to determine equivalent parameter values among different methods. The equivalence is established empirically, on an image-by-image basis using the iterative search described in Section \ref{sec-it-search}. Finally, Section~\ref{sec-clustering} describes our methodology for assessing similarity between operators. For all of our experiments, we use the set of 300 images from the BSDS300 Berkeley Segmentation Dataset and Benchmark \cite{martin2001database}.

\subsection{Edge-preserving smoothing operators}\label{sec-smoothing-operators}

When selecting a set of methods to include in this paper, our aim was not to produce an exhaustive review, but rather assemble a representative set of edge-aware operators which are both diverse in terms of their approaches and prevalent in the literature, as methods to which newly developed filters are commonly compared. We also opted to include methods of varying computational complexity and some of historical significance in the field. Finally, we include operators that are particularly well-suited for specific applications such as texture removal or multiscale image manipulation.

\textbf{Bilateral filtering (BLF).} A classic paper by Tomasi and Manduchi \cite{tomasi1998bilateral} laid the foundation of bilateral filtering, where geometric and photometric closeness between pixels are combined into a single “bilateral” filter.

\textbf{Weighted least squares (WLS).} An image-guided optimization-based approach by Farbman et al. \cite{farbman2008edge} yielding an edge-preserving filter that is well suited for progressive coarsening and multiscale detail manipulation.

\textbf{Multiscale image decomposition based on local extrema (EXT).} A method by Subr et al. \cite{subr2009edge} in which details in an image are identified as oscillations between local extrema. This unique approach to multi-scale image decomposition allows effective smoothing of textures characterized by strong gradients.

\textbf{L0 gradient minimization (L0).} A global optimization approach by Xu et al.~\cite{xu2011image}, which attempts to minimize the discrete number of non-zero gradients in the output.

\textbf{Fast local Laplacian filters (FLLF).} Paris et al.~\cite{paris2011local} use Laplacian pyramids modified to enable edge-aware image processing. Laplacian coefficients are assigned as belonging to either edges or details in the image, based on a user-defined threshold parameter that defines a remapping function used for separation between the two. This operator was later accelerated \cite{aubry2014fast}, which is the version that we have used in this work.

\textbf{Domain transform (DOM).} An approach by Gastal and Oliveira~\cite{gastal2011domain} that enables real-time filtering via an iterative domain transform that adaptively warps the 2D input signal in a series of fast 1D filtering operations.

\textbf{Relative total variation (RTV).} In an algorithm designed for texture removal, Xu et al.~\cite{xu2012structure} define the “inherent variation” and “relative total variation” measures to distinguish texture elements from semantically meaningful structures in an image.

\textbf{Guided image filtering (GIF).} The guided image filter by He et al.~\cite{he2013guided} uses a guidance image to filter the input image with accordance to its structures. To perform edge-preserving filtering, we used the input image as a guidance image for itself, just as the authors did in the original paper.

\textbf{Minimum Spanning Tree (MST).} This filter by Bao et al.~\cite{bao2014tree} introduces the notion of “pixel connectedness” by defining individual pixels as nodes in a minimum spanning tree of the input image. This notion is used alongside with spatial and color/intensity differences between pixels to achieve strong smoothing, while preserving major edges.

\textbf{Fast weighted least squares (FGS).} Another global filtering approach, this work by Min et al.~\cite{min2014fast} offers a computational advantage to the traditional task of optimizing a global objective function consisting of a data constraint and a smoothness prior. The authors approximate the solution to the original linear system by solving a sequence of 1D linear sub-systems.

\textbf{L1 image transform (L1).} Bi et al.~\cite{bi20151} present an image transform based on the $L_1$-norm with the goal of achieving piecewise image flattening. Their method produces nearly piecewise constant images by minimizing an energy function given by a weighted sum of three components, describing local flattening, global sparsity and image approximation terms.

\subsection{Primary controlling parameters}\label{sec-primary-parameters}

Smoothing operators feature at least one primary controlling parameter, which specifies the desired degree of smoothing; increasing its value yields progressively strongly smoothed outputs. In many cases, there are one or more additional parameters that further tune the method's behavior. Defining a mapping between sets of parameters whose number and goals differ from one method to another is extremely challenging, so for our analysis we identify and focus on a single primary controlling parameter for each operator. For each method's primary parameter, we identify its maximal value as the value above which either no visible change occurs, or none of the original image structures are detectable in the output. These maximal values were determined empirically using the images from the BSDS300 dataset. In our analysis, all primary parameters have a minimal value of zero, which corresponds to the parameter that returns the unfiltered input image. Table \ref{parameters-table} lists the primary parameters for the methods listed earlier, alongside with their corresponding maximal values. In cases where methods require additional input parameters, their default values are used.

\begin{table}
\caption{Primary controlling parameters and their maximal values for the operators listed in Section \ref{sec-smoothing-operators}.
}
\centering
\label{parameters-table}
\begin{tabular}{|p{1cm}|p{1.1cm}|p{4.1cm}|p{0.5cm}|}
\hline
\textbf{Method}&\textbf{Param}&\textbf{Description}&\textbf{Max}\\
\hline
\hline
BLF & $\sigma_r$ & $\sigma_d$ spatial distance, $\sigma_r$ intensity range. We set $\sigma_d = 20\sigma_r$. & 0.5 \\
WLS & $\lambda$ & degree of smoothing & 10\\
EXT & $k$ & width of neighborhood for identification of local extrema & 30\\
L0  & $\lambda$ & degree of smoothing &0.3\\
FLLF & $\sigma_{r}$ & edge-detail separation threshold &10\\
DOM & $\sigma_{r}$ & range standard deviation &5\\
RTV & $\lambda$ & degree of smoothing &0.05\\
GIF & $r$ & local filter window radius &10\\
MST & $\sigma$ & tree distance parameter &3\\
FGS & $\sigma$ & range standard deviation &0.1\\
L1  & $\alpha$ & local sparseness weight &20\\
\hline
\end{tabular}

\end{table}

\subsection{Image attributes}\label{sec-img-attributes}

Linear image processing filters are typically analyzed and compared to each other in the frequency domain. However, this approach is not applicable to spatially variant edge-preserving filters, since preserved local image gradients have a global effect in the frequency domain. Instead, in this work we examine directly the effect that operators have on the magnitudes of image gradients in the spatial domain. Since the specific gradients attenuated by each method, and the degree to which they are attenuated, differ from one method to another, we examine the overall attenuation effect over different image regions.

More formally, let $I$ denote an input image and $J$ its smoothed version.
We denote by $|\nabla I_p|$ and by $|\nabla J_p|$ the magnitude of the gradient vector at pixel $p$ before and after applying the operator, respectively. We measure the ratio between the sum of gradient magnitudes in the smoothed image over the corresponding sum in the original image. We denote this ratio for the entire image by $SO$ (Smoothed/Original):

\begin{equation}
SO = \frac{\sum_{p \in \mathcal{P}} |\nabla J_p|}{\sum_{p \in \mathcal{P}} |\nabla I_p |}.
\end{equation}
The $SO$ attribute reflects the attenuation of gradient magnitudes across the full set of image pixels $\mathcal{P}$. Since our goal is to analyze edge-preserving filters, we are particularly interested in measuring the attenuation of gradients in areas of the image that do not contain salient edges \emph{separately} from areas in the vicinity of such edges.
This distinction allows us to assess how different operators resolve their two contrasting goals of smoothing out the image while simultaneously preserving significant gradients.

In order to separately assess these two contradicting effects, we construct a \emph{smooth mask} $\mathcal{M}_S$ as described below, and define an \emph{edge mask} as its complement:
$\mathcal{M}_E = \mathcal{P} \setminus \mathcal{M}_S$. We then compute the $SO$ attribute over the pixels in $\mathcal{M}_S$ and, separately, over the pixels in $\mathcal{M}_E$, resulting in two image attributes that we refer to as $SO_S$ and $SO_E$, respectively.

\begin{figure}[htb]
\centering
\includegraphics[width=.5\textwidth,keepaspectratio]{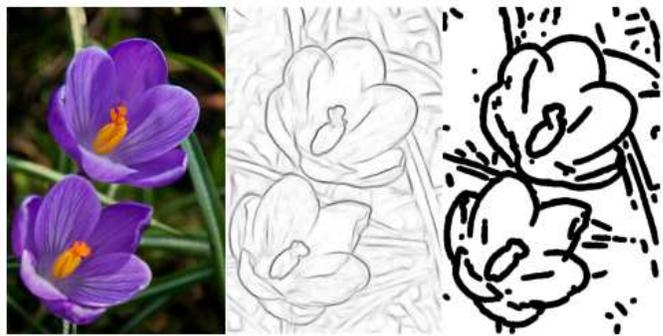}
\caption{\label{fig:masks}
Input image (left), inverted edge map (middle), and the smooth mask (right).}
\end{figure}

In order to construct the smooth mask we first compute an \emph{edge map}, which specifies for each pixel the likelihood that it belongs to a salient edge in the image. We use the structured forests edge detection method \cite{dollar2013structured} for this purpose, but other edge detection algorithms could be used instead. The smooth mask $\mathcal{M}_S$ is then constructed by thresholding the edge map, to remove all pixels whose edge saliency is greater than 0.3, and eroding the resulting binary image using a disk with a 5-pixel radius. The erosion ensures a safe distance from the salient edges in the image. This process is illustrated in Figure \ref{fig:masks}.

We are also interested in examining the effect of different operators on image attributes that have been shown to play an important role in subjective perceptual studies conducted on the assessment of TMOs. In particular, changes in overall brightness and color were shown to significantly affect image quality perception \cite{vcadik2006image}. We measure these changes between $I$ and $J$ in the perceptually uniform CIE $Lab$ color space.

The overall change in perceived image brightness is measured as the average of the ratios between the processed and the original pixels' $L$ channel:
\begin{equation}
\Delta L = {1 \over |\mathcal{P}|}\sum_{p \in \mathcal{P}} \frac{L(J_p)}{L(I_p)}.
\end{equation}
Changes in color are measured as the sum of the differences in the chroma channels $a, b$, averaged over all pixels:
\begin{equation}
\Delta C = {1 \over |\mathcal{P}|}\sum_{p \in \mathcal{P}} \sqrt{(a(I_p) - a(J_p))^2 + (b(I_p) - b(J_p))^2}.
\end{equation}

Image contrast is yet another feature that was found to be of importance in subjective studies of tonemapped images. Matkovic et al. \cite{matkovic2005global} use a weighted average of local contrasts at various spatial frequencies to define a global contrast factor (GCF). In their study, the weighting function is determined by a best fit to “perceived contrasts” that were obtained from a user study in which participants were asked to rank a series of images according to their contrasts. In this work we define the change in image contrast as ratio of GCF between the smoothed and input images.

\subsection{Empirical parameter equivalency}\label{sec-it-search}

While most methods can achieve smoothing effects varying from mild to strong, their behavior across this range may differ. In our methodology, the results of smoothing an input image by different algorithms are directly comparable to one another, only if they represent an equivalent smoothing level, as reflected by their corresponding $SO$ scores.

More specifically, we define the smoothing level as $1 - SO$. Since $SO \in [0,1]$, the original unfiltered image would have a smoothing level of $0$, while an image where most of the gradients have been nearly eliminated would have a smoothing level of $1$. 
Thus, given an input image and a target smoothing level $SL \in [0,1]$, we search the parameter range of each operator for the value that produces a result whose $SO$ score most closely matches $1 - SL$.

Since $SO$ decreases monotonically with the increase in the value of the controlling parameter (in all the methods we have experimented with), the search can be carried out using a procedure similar to maximum-seeking or hill-climbing approach \cite{shubert1972sequential}. Specifically, we iteratively narrow down the searchable parameter range by continuously moving in the direction of an upward gradient (increasing similarity of the smoothing level). The initial parameter range is defined by 0 and the maximal effective parameter value, as defined separately for each method (see Table \ref{parameters-table}). The search terminates when one of the following conditions is met:

\begin{itemize}
\item The obtained smoothing level is within $10^{-3}$ from the target smoothing level.
\item The absolute difference between smoothing levels on the leftmost and rightmost points of the current iteration interval is less than a small predefined threshold.
\end{itemize}

\subsection{Similarity of smoothing operators}\label{sec-clustering}

Once equivalency of parameters between different methods is empirically established as described above (on a per-image basis), it becomes possible to quantitatively measure the similarity of their results. To this end, we employ the structural similarity index (SSIM) \cite{wang2004image}.
Operators that produce visually similar images result in higher SSIM scores, as they are able to achieve a comparable effect not only in terms of the overall smoothing level they achieve, but also in terms of perceptually important changes in structural information. This approach allows us to cluster smoothing operators and deduce the distinct groups of methods that behave similarly to each other.
\section{Results}\label{sec-results}

In this section we apply the methodology described in the previous section to the methods listed in Section \ref{sec-smoothing-operators}.
We begin by showing that these operators are characterized by unique profiles of progressive gradient elimination (\ref{sec-prog-grad-elimination}). In order to perform a comparative analysis of operators, we establish a common baseline of target smoothing levels and use it to compare the effect different filters have on three perceptually important image features: overall brightness, color, and contrast (\ref{sec-perceptual-features}). Section \ref{sec-feature-comb} demonstrates a joint analysis of smoothing vs. edge-preservation, which provides insights on edge-aware operators' behavior. Section \ref{sec-results-clustering} demonstrates our methodology for clustering the operators according to behavioral similarity.

\subsection{Gradient elimination profiles}\label{sec-prog-grad-elimination}

Normally, increased smoothing levels correlate with elimination of increasingly strong gradients in the image. In order to assess the progressive smoothing effect of an algorithm, each pixel in an input image is assigned to one of one hundred bins, based on its original gradient magnitude. Low-numbered bins contain pixels with lower original gradient magnitudes, while high-numbered bins contain pixels with stronger ones (likely belonging to edges in the original image).

\begin{figure}[tb]
\centering
\includegraphics[trim={0.01cm 0 0 0},clip,width=\linewidth]{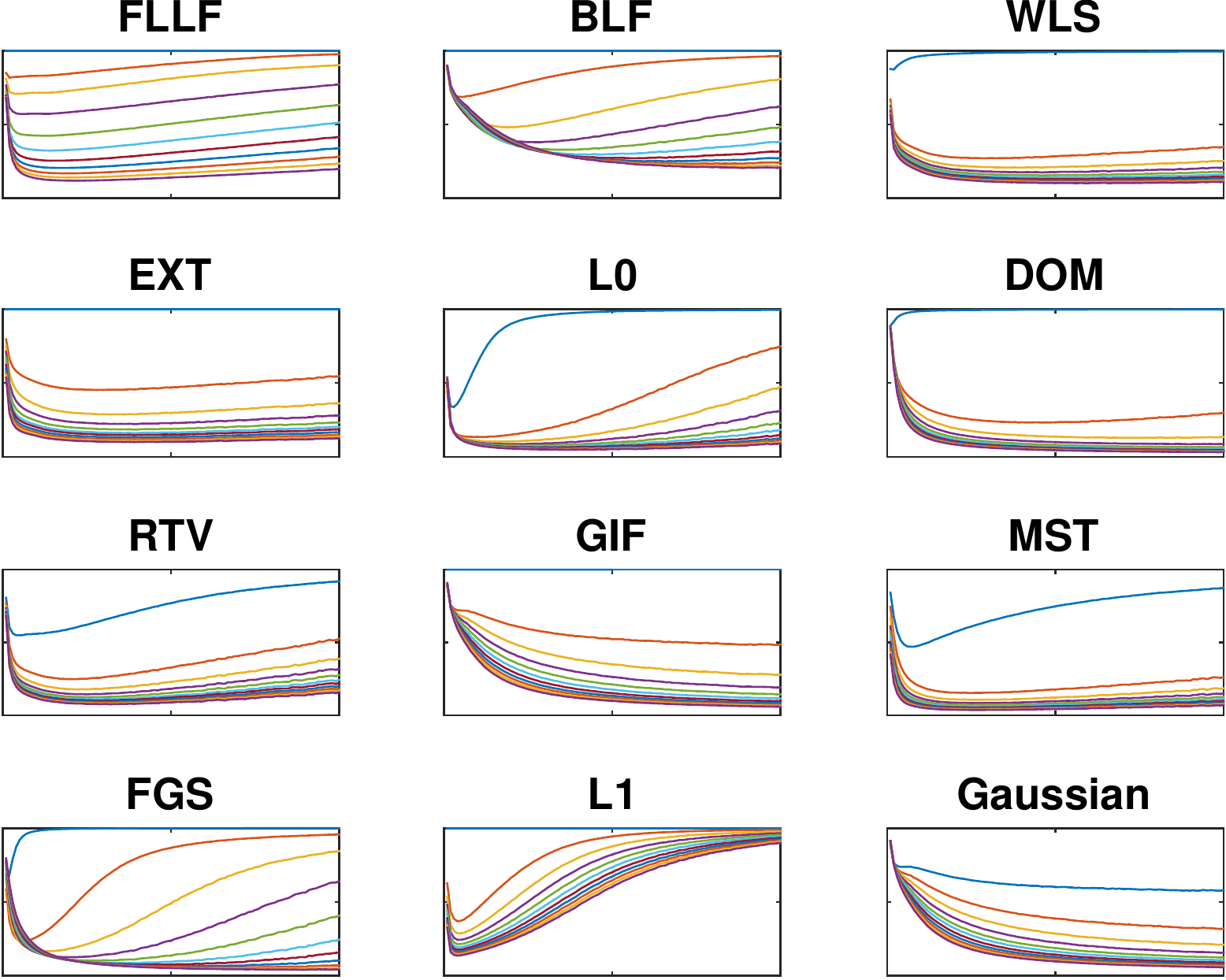}
\caption{\label{fig:progressive-gradient-elimination}
Progressive gradient elimination profiles. The horizontal axis corresponds to gradient bins with the magnitude of gradients increasing from left to right. The vertical axis is the $SO$ value computed over the pixels in each bin, averaged over all images in BSDS300. There are 11 curves for each method, corresponding to a uniform sampling of each method's parameter range.
}
\end{figure}

Figure \ref{fig:progressive-gradient-elimination} plots the aforementioned progressive gradient elimination profiles for each method, by reporting the $SO$ attribute evaluated over the pixels in each bin, averaged over all images in the BSDS300 dataset. Each curve corresponds to a particular primary parameter value: as these values increase, stronger gradients diminish in magnitude and result in lower curves. There are 11 curves for each method, corresponding to a uniform sampling of that method's parameter range.

There are several insights that can be gained from examining such profiles. Firstly, they clearly show how differently methods treat weak and strong gradients: in some methods, such as L0 and L1, the difference in $SO$ values between low- and high-numbered bins is maintained large across the parameter range (steep curves), while in others, such as WLS and MST, the same trend, while evident, is much less pronounced. Methods that differ significantly in their effect on weak and strong gradients tend to produce more piecewise-constant outputs, as they are able to aggressively smooth out low-gradient areas in the image while at the same time preserving stronger, structure-defining gradients. Secondly, the varying spacing of curves in each graph reveals how sensitive each method is to changes in the parameter value. For example, in WLS, DOM and MST, an increase in parameter value causes a stronger change in gradient attenuation when done at the lower end of the parameter range (top curves) than an equal increase applied at the higher end of the range (bottom curves). In contrast, for FLLF the curves are much more evenly distributed across the entire range of parameter values.

\subsection{A common baseline of smoothing levels}
\label{sec-param-equiv}

\begin{figure}
\centering
\includegraphics[trim={0.1cm 0.1cm 0.1cm 0.05cm},clip,height=0.185\textheight,keepaspectratio]{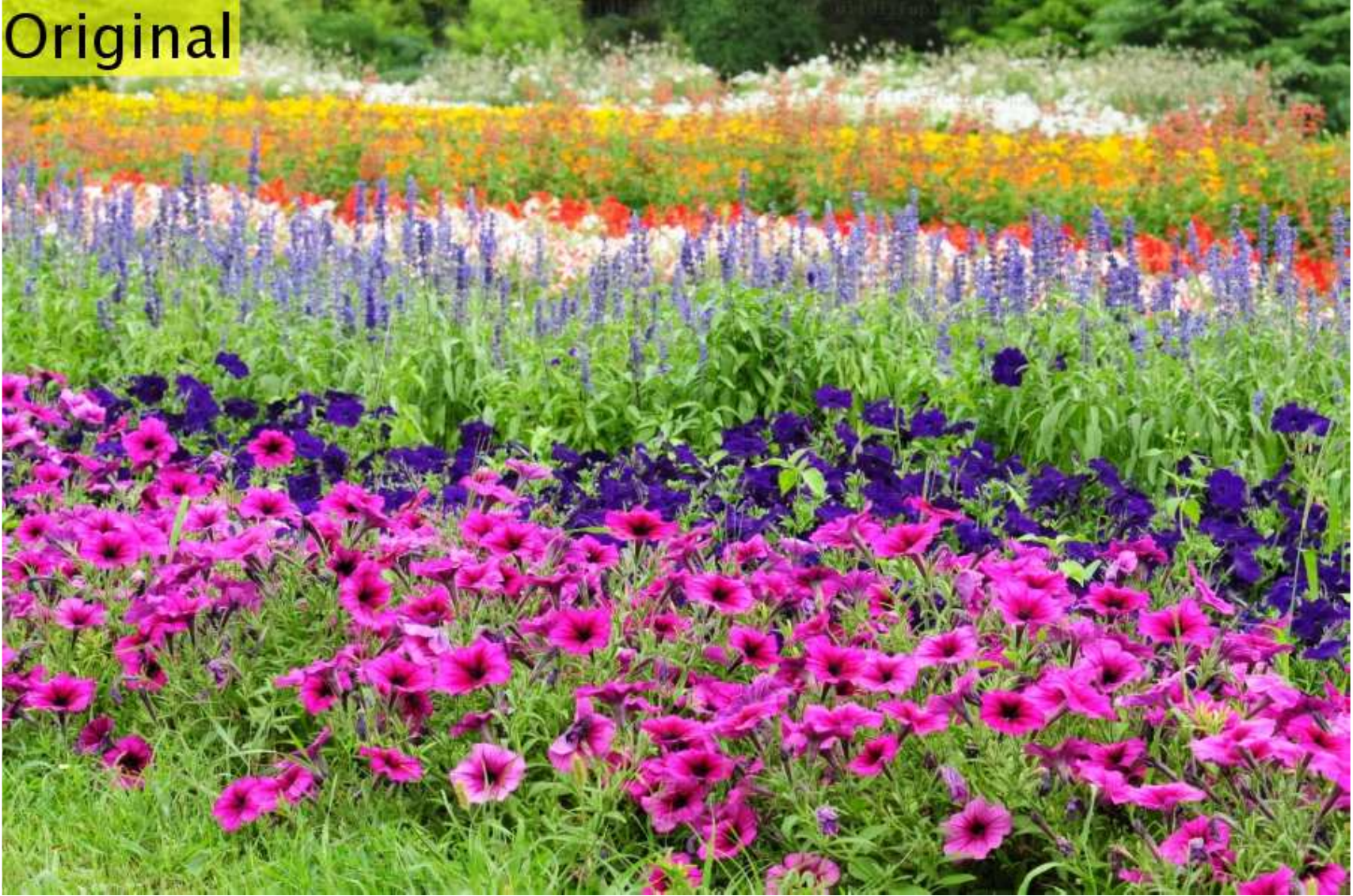}
\includegraphics[trim={0.1cm 0.1cm 0.1cm 0.05cm},clip,height=0.185\textheight,keepaspectratio]{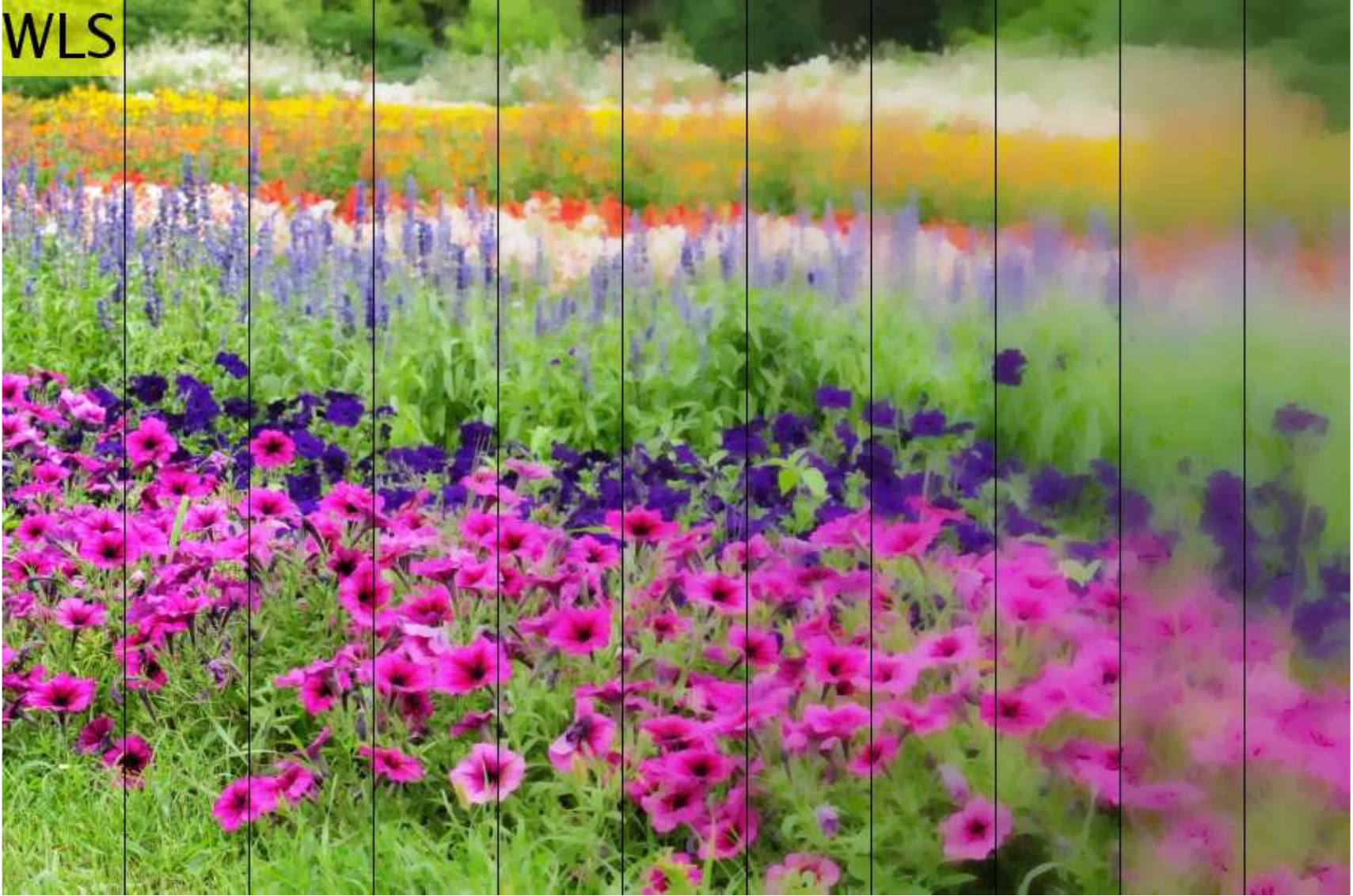}
\includegraphics[trim={0.1cm 0.1cm 0.1cm 0.05cm},clip,height=0.185\textheight,keepaspectratio]{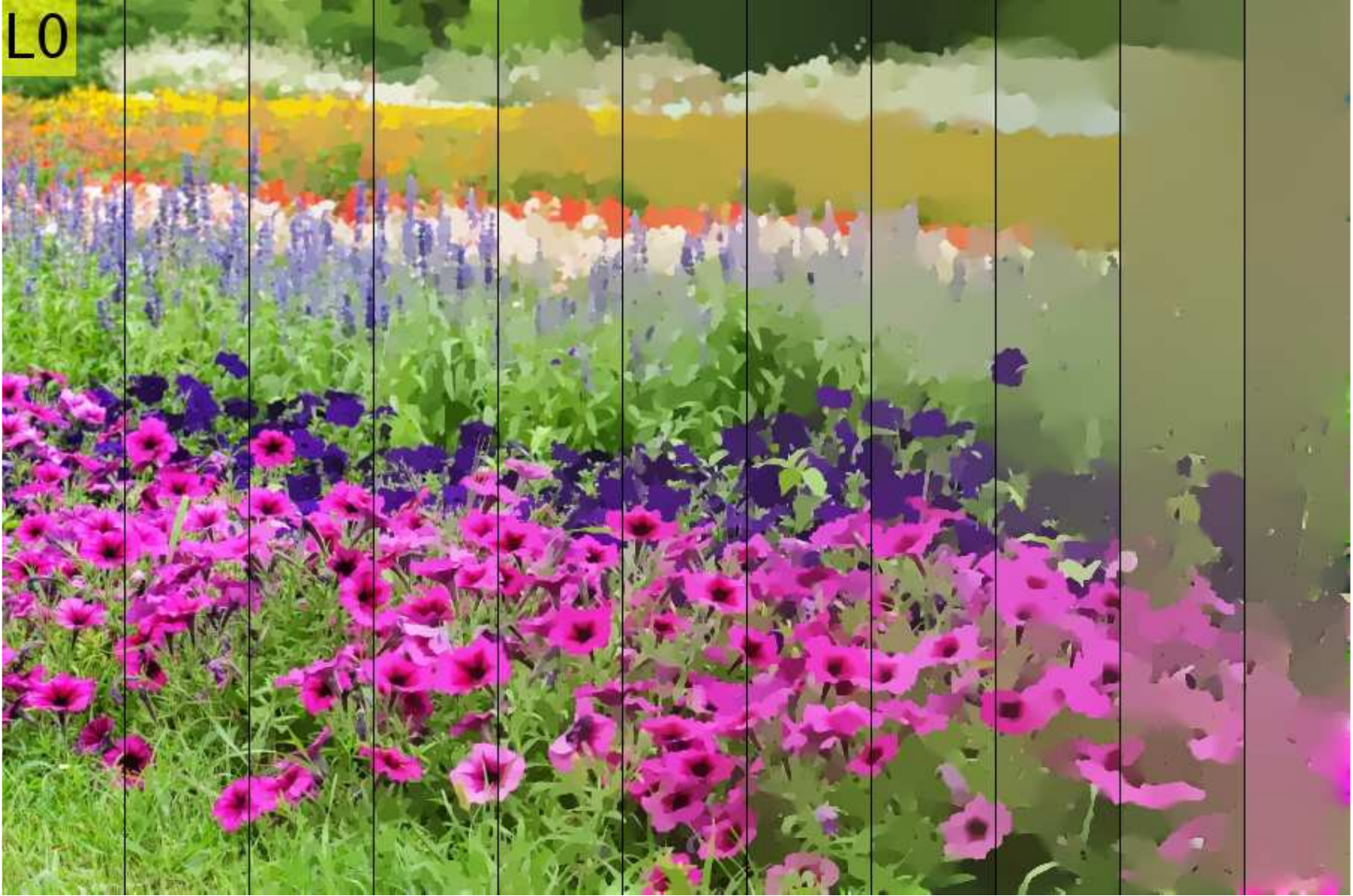}
\includegraphics[trim={0.1cm 0.1cm 0.1cm 0.05cm},clip,height=0.185\textheight,keepaspectratio]{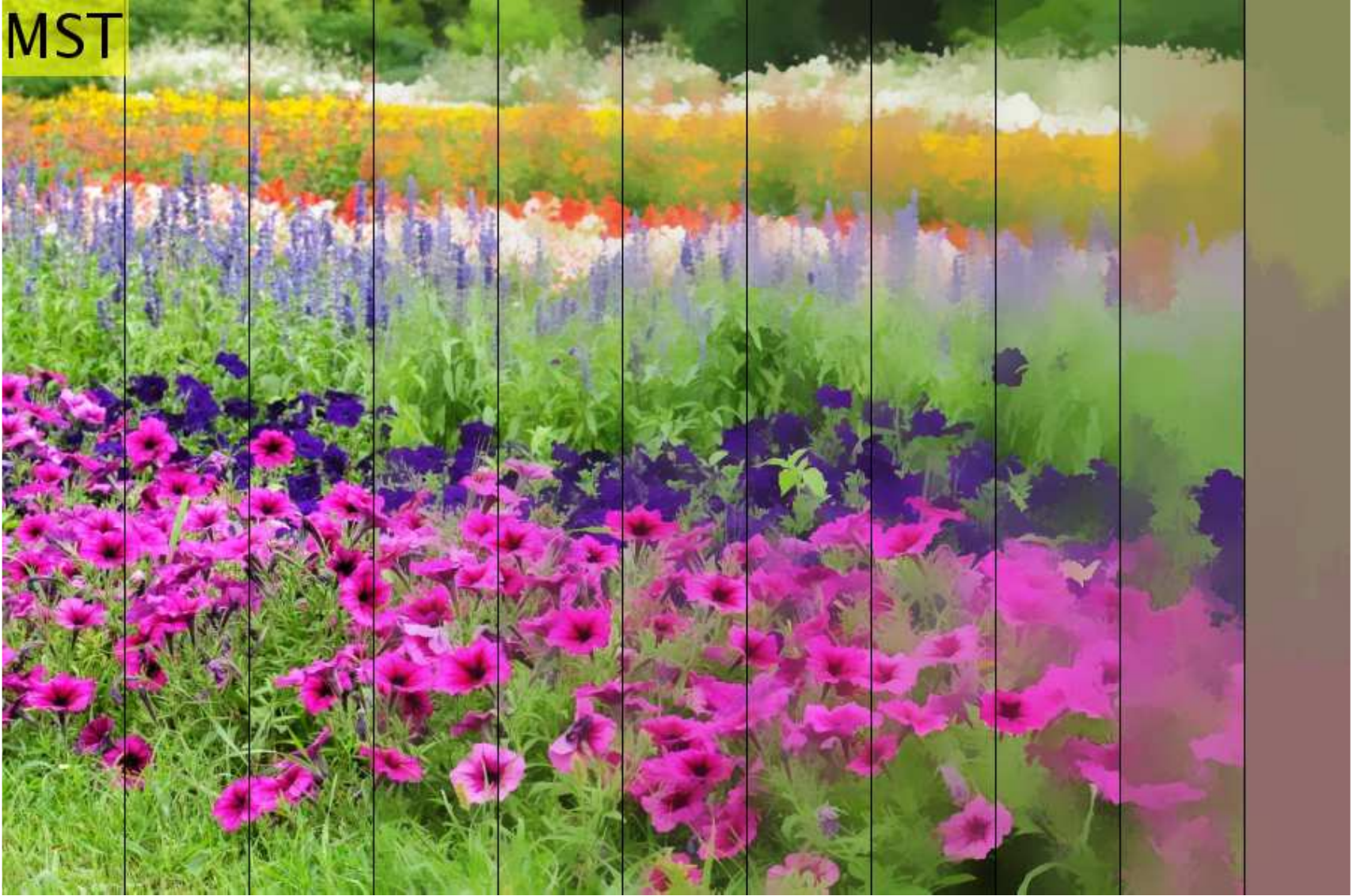}
\includegraphics[trim={0.15cm 0.15cm 0.15cm 0.05cm},clip,height=0.185\textheight,keepaspectratio]{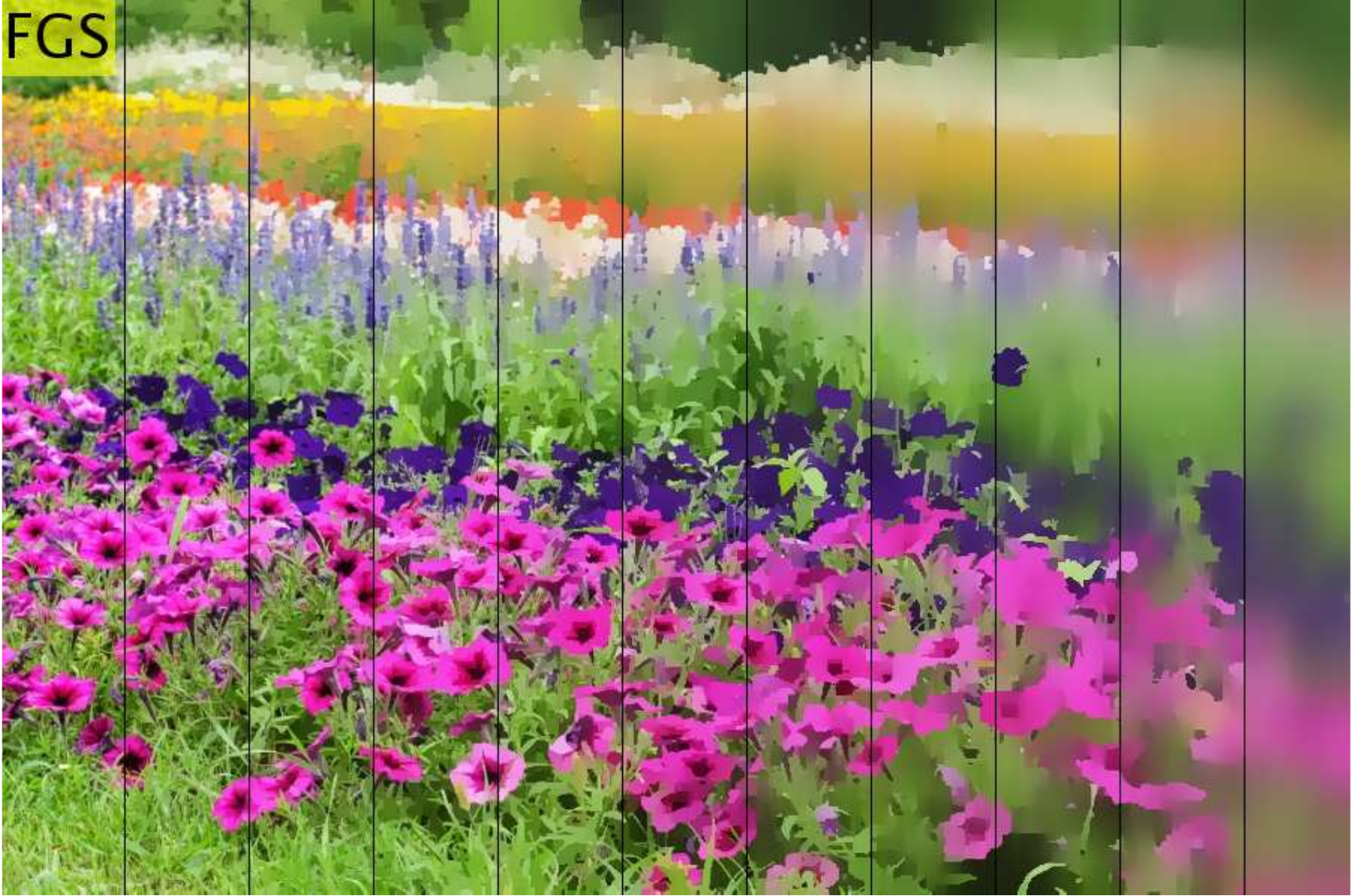}
\includegraphics[height=0.02\textheight,width=0.38\textwidth]{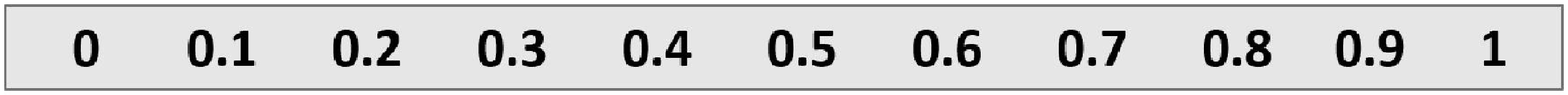}
\caption{\label{fig:Equivalent smoothing levels}
Equivalent smoothing levels as obtained by several different methods. The original image is on top, and the target smoothing level inside each vertical strip is shown at the bottom.
}
\end{figure}

In Section \ref{sec-it-search} we have described how to determine equivalent parameters across different methods by matching a target smoothing level for a specific image. In order to compare different methods we use a common baseline consisting of ten target smoothing layers ranging from 0.1 to 1.
Figure \ref{fig:Equivalent smoothing levels} illustrates the corresponding results obtained by four different methods. Each image is split into eleven vertical slices, with increasing smoothing levels from left to right. The leftmost slice is the original input image, and the rightmost slice represents the approximation of smoothing level 1, i.e., the strongest smoothing that the method is capable of achieving.
The different look produced by each operator is readily apparent in this visual comparison, particularly at the higher smoothing levels.

\begin{figure}[tb]
\centering
\includegraphics[width=1.0\linewidth]{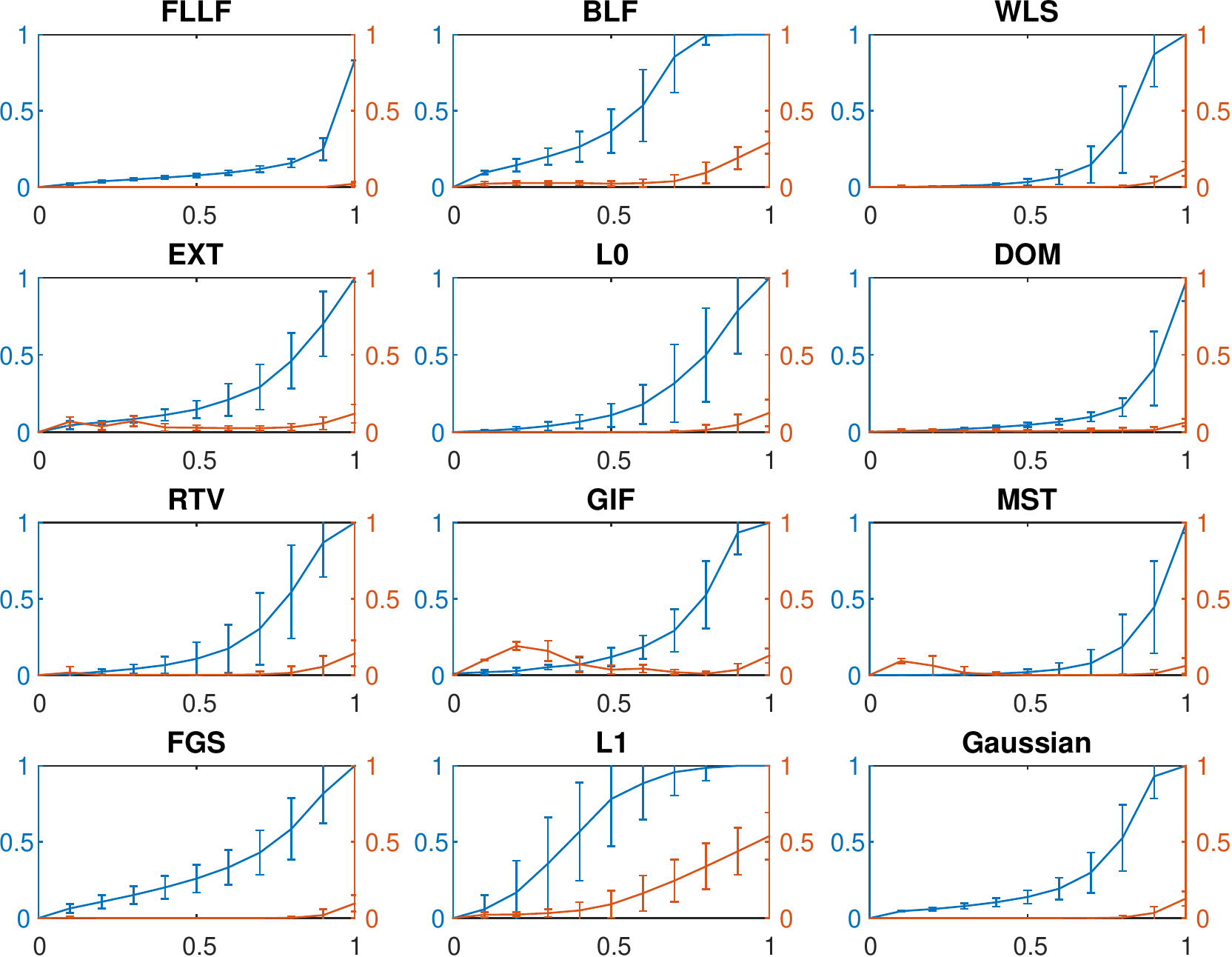}
\caption{\label{fig:Parameter equivalency}
Normalized parameter values required to reach smoothing levels in [0, 1] range ($x$-axis) are plotted in blue for different methods (left $y$-axis). Match accuracy, i.e., the difference from the target smoothing level is plotted in orange and uses the right $y$-axis.}
\end{figure}

The blue plots in Figure \ref{fig:Parameter equivalency} show the parameter values of different methods that were found to match the ten smoothing levels. The blue curve is the average parameter value (normalized to the $[0,1]$ range) over all the images in BSDS300. The error bars show the variance of the parameter value at each smoothing level across this dataset, which generally tends to increase for higher smoothing levels. Figure \ref{fig:Parameter equivalency} also plots in orange the difference between the target smoothing level and the closest matching level that each operator was able to achieve. This shows how well each operator is able to match the target smoothing level across the range, as not all operators are flexible enough to accurately match all levels.

From these plots one can determine the effective parameter range of each method by examining the values that span the possible smoothing levels. Areas where methods do not achieve a close approximation to a given smoothing level can also be detected. For example, we see that MST deviates from the target at mild smoothing levels, but matches strong smoothing levels well: this corresponds to a smoothing behavior characterized by harsher gradient diminishing behavior that is less receptive to fine-tuning by the primary controlling parameter. Few methods are able to approximate complete gradient elimination, as is evident by slight uplifts in the orange precision curves as the smoothing levels approach 1. Finally, the non-linear response to increasing parameter values is reflected in the shapes of the blue parameter curves, which typically become more steep in the region of the higher smoothing levels: methods seem to be more sensitive to small changes in parameter values at the lower smoothing levels than they are at higher ones. This is the same phenomenon that was evident in the non-uniform spacing of the curves in Figure \ref{fig:progressive-gradient-elimination}.

\begin{figure*}
	\begin{tabular}{@{}ccc@{}}
		\includegraphics[trim={0.03cm 0 0 0},clip,width=.33\textwidth]{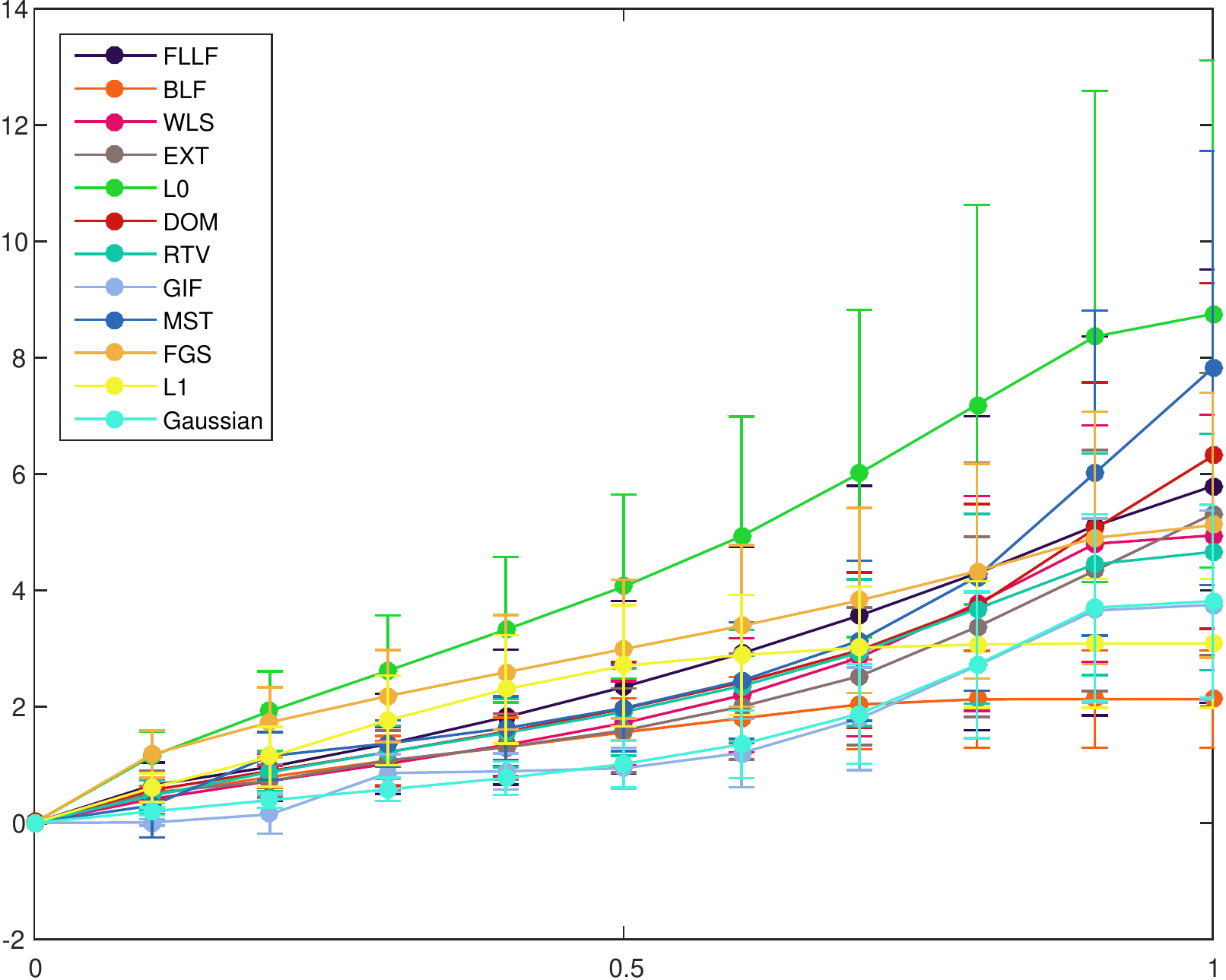} &
		\includegraphics[trim={0.03cm 0 0 0},clip,width=.33\textwidth]{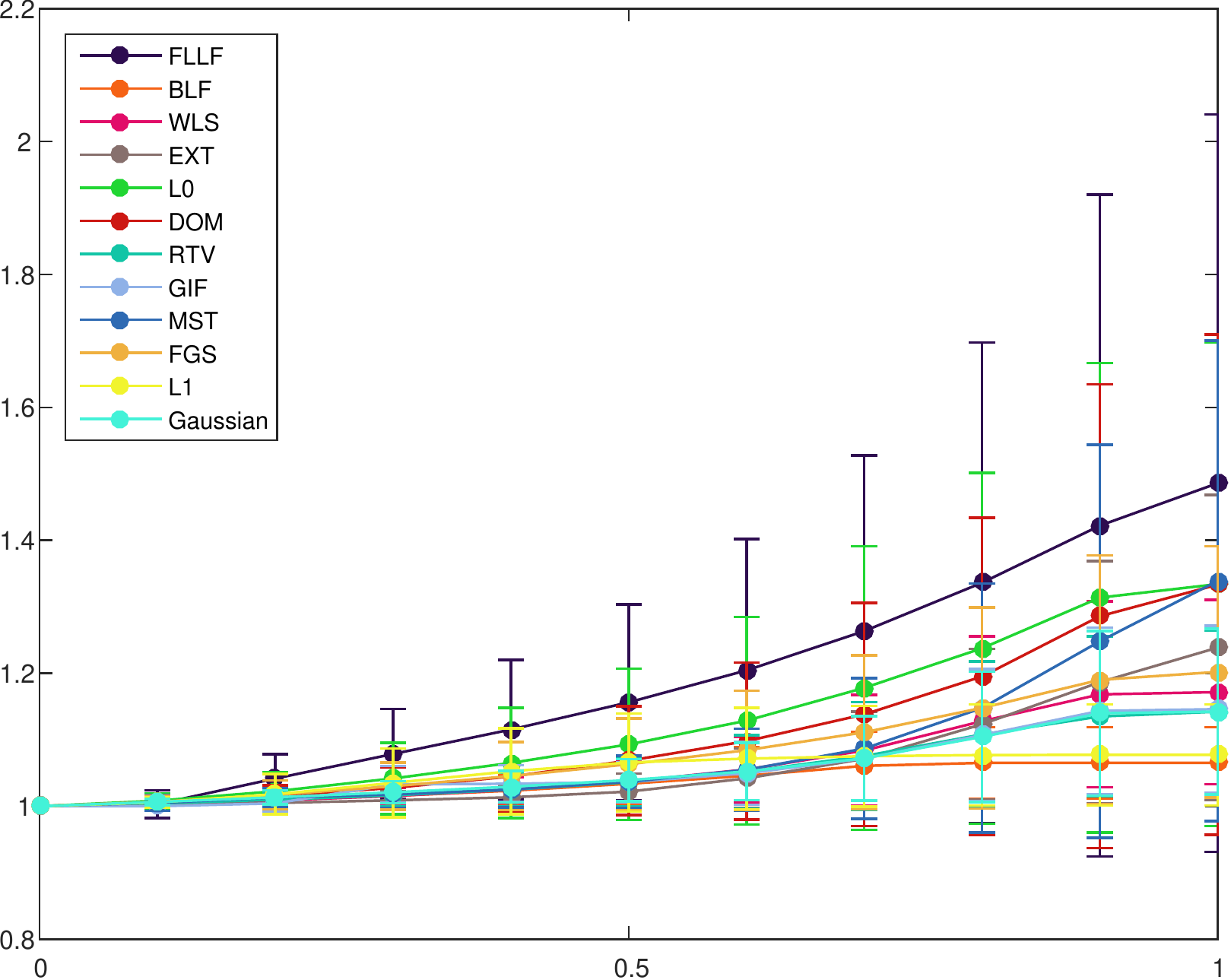} &
		\includegraphics[trim={0.03cm 0 0 0},clip,width=.33\textwidth]{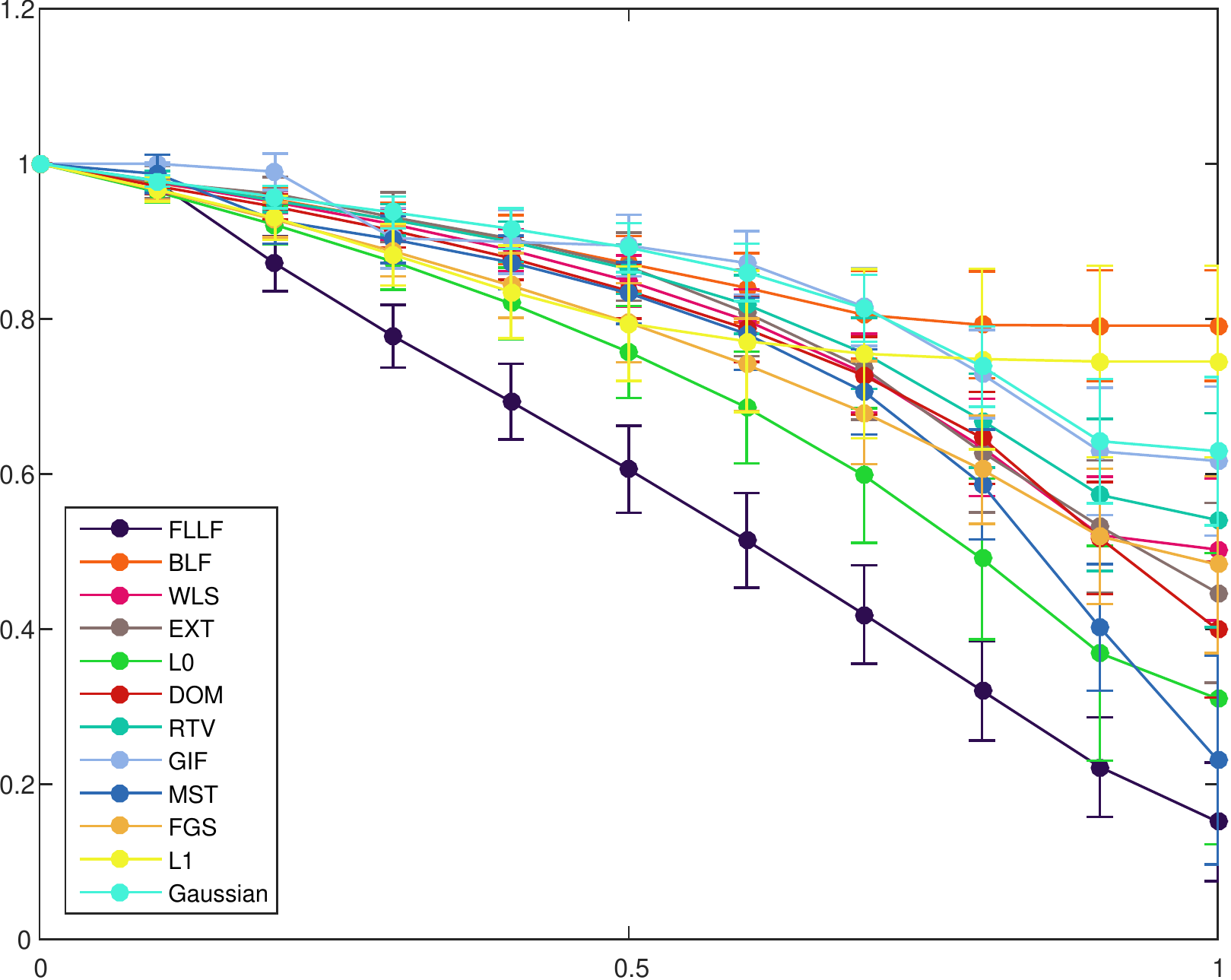}
	\end{tabular}
	
	\caption{\label{fig:CCB}
		Smoothing effect of different operators on image attributes across the smoothing level range: color (left), brightness (middle), and contrast (right).
	}
\end{figure*}

\subsection{Effect on perceptually significant attributes}
\label{sec-perceptual-features}

Subjective studies have indicated that overall image quality, as perceived by the human visual system, depends on several perceptual image attributes that include brightness, contrast and color. Consequently, changes in these attributes strongly affect how the smoothed image is perceived. Edge-preserving operators cannot, of course, be expected to perfectly maintain original input attributes, but methods do differ in the changes that occur in these attributes as a result of smoothing operations. In terms of image quality, one might favor an algorithm that preserves original colors even at high smoothing levels, e.g., in tasks such as image abstraction. Alternatively, a user might opt for the method which least affects image contrast. 

In the previous section we demonstrated that the range of possible smoothing levels provides an objective baseline for comparison between various operators. Figure \ref{fig:CCB} uses this baseline to compare the different methods based on their effect on aforementioned image attributes, namely overall brightness, color, and contrast. This figure demonstrates that, in general, the evaluated operators tend to produce images characterized by loss in image contrast and increased brightness values, in addition to deviations from the original colors. However, differences may be seen in the extent to which these attributes are altered, as well as in their dependency on the input image.

Figure \ref{fig:CCB} also illustrates that the differences between methods may vary across the smoothing levels; for example, some methods, but not others, are characterized by only mild changes in features up to a certain smoothing level, but spike at higher levels (see MST effect on color). At the same time, methods such as L0 tend to more strongly affect original image colors across the entire smoothing levels range.

\begin{figure}[htb]
	\centering
	\includegraphics[trim={0.03cm 0 0 0},clip,width=0.8\linewidth]{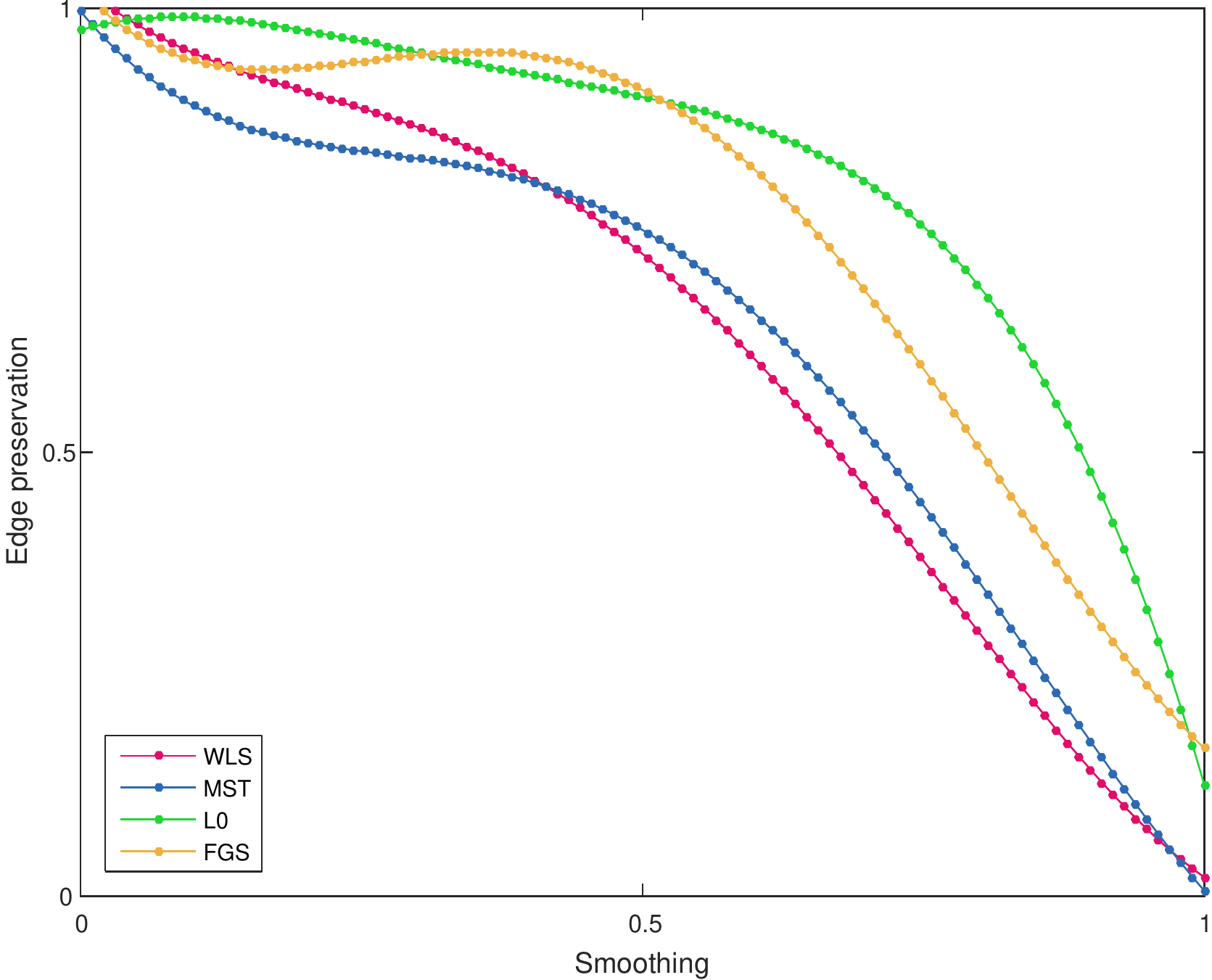}
	\caption{\label{fig:Smooth vs. Edges}
		Edge-preservation as a function of smoothing for different operators.
	}
\end{figure}

\subsection{Smoothing vs. edge-preservation}\label{sec-feature-comb}

When applying edge-preserving smoothing, the goal is, on the one hand, to achieve the desired amount of smoothing in certain areas of the image, while, on the other hand, retaining the salient edges separating between them. With our methodology one may examine the interplay between these two goals by plotting the joint behavior of the $SO_S$ and $SO_E$ attributes of each of the evaluated methods. Figure \ref{fig:Smooth vs. Edges} plots a polynomial fit of the typical range of combinations for four of the evaluated methods. The more concave curves of the L0 and FGS operators indicate that they are capable of achieving a higher level of smoothing in the smooth mask regions $\mathcal{M}_S$, with less gradient attenuation near salient edges (under the $\mathcal{M}_E$ mask). In contrast, WLS and MST cannot achieve such combinations and are characterized by less convex curves.

\begin{figure}
\centering
\includegraphics[width=\linewidth]{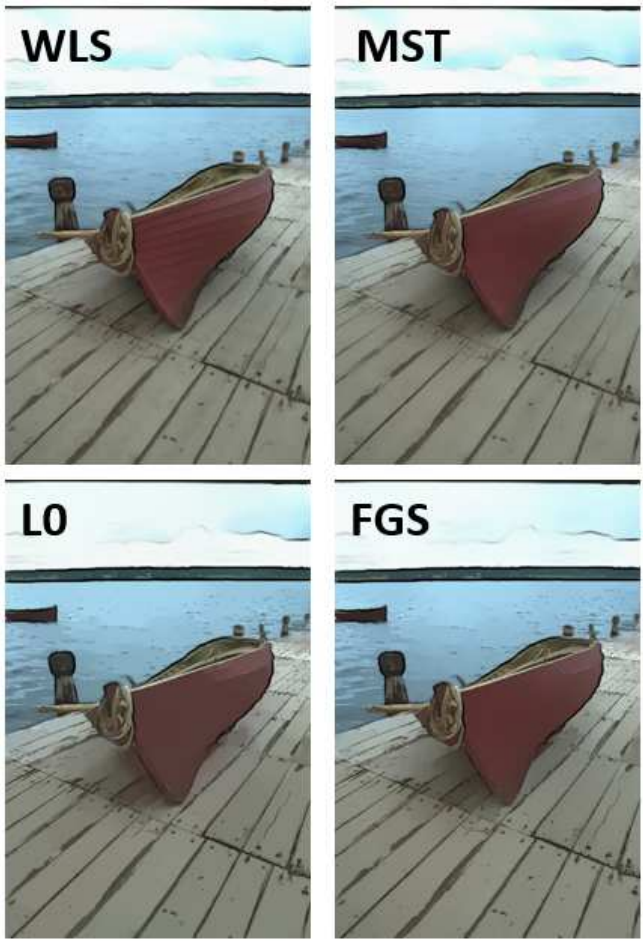}
\caption{\label{fig:Abstraction}
Abstraction on equivalently smoothed images (50\% gradient reduction). WLS and MST are less uniform inside the smoothed regions than L0 and FGS.
}
\end{figure}

\begin{figure}[htb]
\centering
\includegraphics[width=\linewidth]{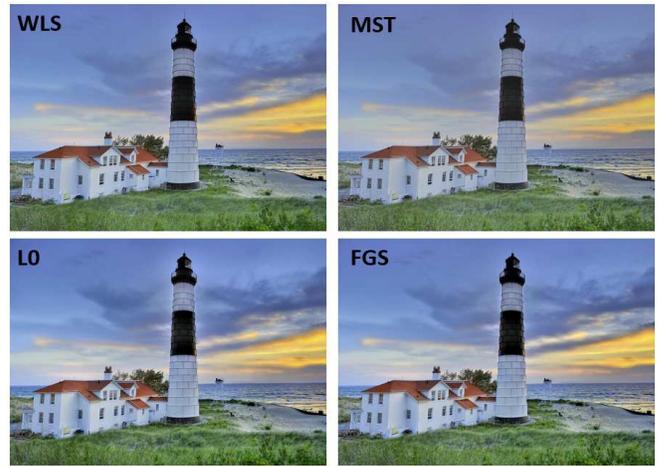}
\caption{\label{fig:DetailEnhancement}
Detail enhancement using equivalently smoothed images (50\% gradient reduction). L0 and FGS exhibit some darkening in the grass texture, and have stronger halos.
}
\end{figure}

Such observations on the typical behavior of operators can help us better understand the tasks for which they are best suited for. To illustrate this idea, we look at two common tasks involving edge-preserving smoothing. In image abstraction, details are removed to create a cartoon-like result, which is preferably characterized by regions that are uniform in color. In detail enhancement, the image is separated into a base and detail layers, with the latter undergoing some form of boosting before being recombined with the base layer. In this case, the smoothing operation benefits from finer removal of details than that required for abstraction. Thus, methods such as L0 and FGS, would be better suited for image abstraction, while WLS and MST might be a better choice for detail enhancement. We provide examples of outputs of these four methods for both tasks in Figures \ref{fig:Abstraction} and \ref{fig:DetailEnhancement}. Note that L0 and FGS are indeed able to achieve more uniform smoothing, resulting in more piecewise-constant cartoon-like results, while WLS and MST can produce crisp details, with noticeably less darkening of regions.

\subsection{Clustering of operators}\label{sec-results-clustering}

As mentioned earlier, it is important to identify similarity between methods for a number of reasons. Briefly, it can aid developers of edge-preserving operators in choosing a set of methods to which to compare their new algorithms. For filters designed with specific tasks in mind, such as tonemapping or texture removal, the best choice would be to compare them to other methods similar in nature in order to bring to light task-specific benefits of the new algorithm. For more general-purpose operators, a comparison to a more diverse set of algorithms would be preferred, both to indicate which tasks the operator might be best suited for, and to see how its behavior compares to that of different classes of edge-preserving filters. Since users cannot easily distinguish between the different numerical approaches, output styles and behaviors of the methods used, an automated grouping of methods would be beneficial. 

\begin{figure}[htb]
	\centering
	\includegraphics[trim={0.01cm 0 0 0},clip,width=\linewidth]{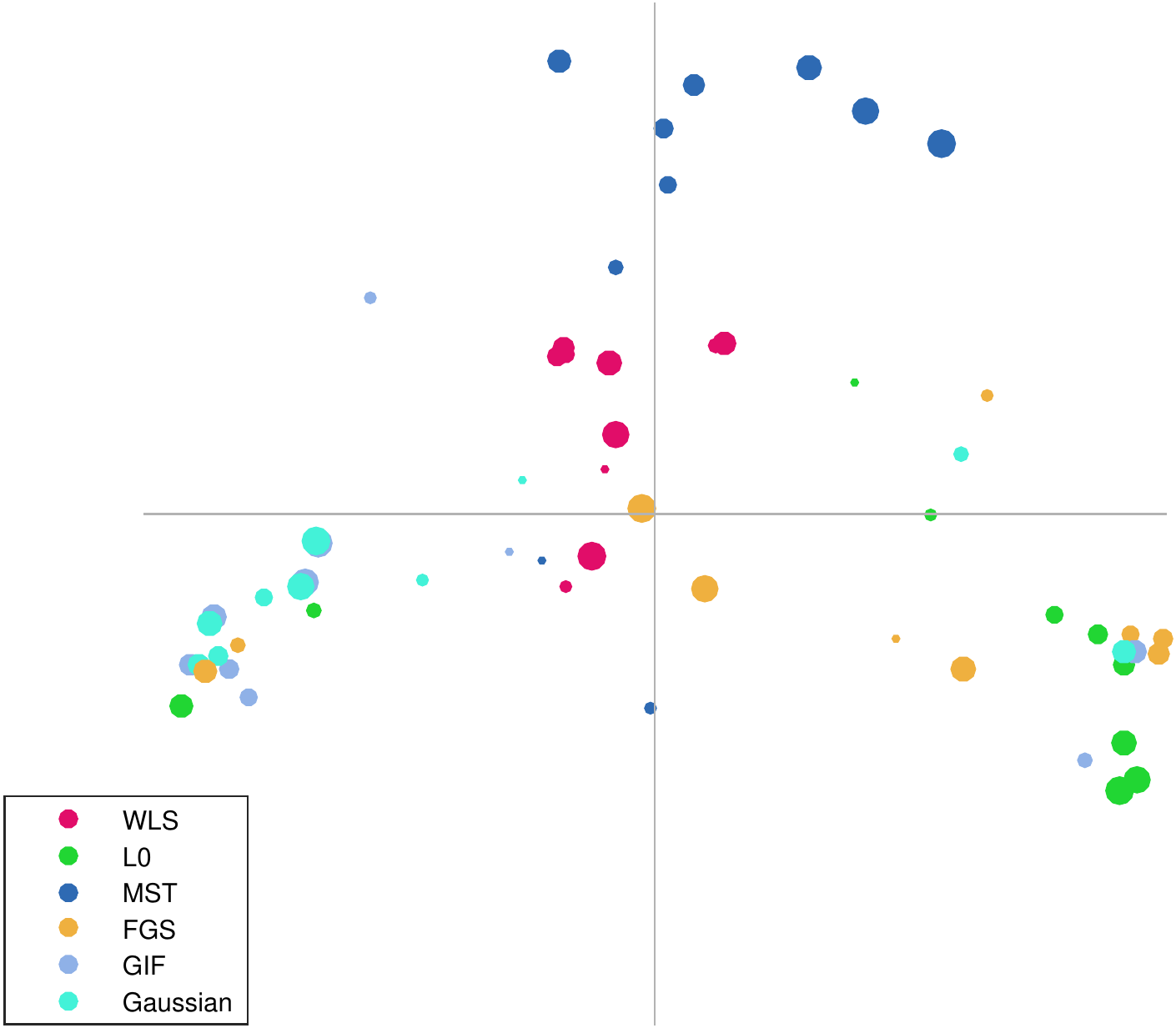}
	\caption{\label{fig:Clustering} Planar embedding of six smoothing operators, using SSIM to measure similarity between equivalently smoothed images.}
\end{figure}

Figure \ref{fig:Clustering} shows the embedding of results of six visually pairwise-similar operators in the plane. The embedding is obtained by  multidimensional scaling (MDS) performed on a distance matrix of average SSIM scores between equivalently smoothed images. Since SSIM scores vary at different smoothing levels, we use increased circle sizes to indicate higher smoothing levels. Figure \ref{fig:Clustering} shows that methods similar in visual style tend to be located in proximity to one another: WLS to MST, L0 to FGS, GIF to Gaussian.

\subsection{Equivalent parameters for commonly used test images}
\label{sec-common-imgs}

As can be inferred from Figure \ref{fig:Parameter equivalency}, finding an input-independent parameter mapping between methods is a challenging task, as evident from the large standard deviation values of the parameter matching curves. For this reason, we cannot offer the reader a set of equivalent parameters \emph{a priori}, i.e., one that would be applicable to any input image. Nevertheless, it is possible to compute such a set for any given image; we therefore plan to make available an interactive user interface, through which users will be able to run our search algorithm to determine equivalent smoothing parameters for any image and target smoothing level using their methods of choice.
\section{Summary and Discussion}
\label{sec-discussion}
\label{sec-conclusions}

The abundance and diversity of edge-preserving filters make it difficult for users to choose which operators might be best suited for their particular needs. Unfamiliarity with implementation details of smoothing operators may hinder users not only in the selection of any particular method, but also when choosing the parameters to be passed on to the chosen algorithm. In this paper, we offer ideas on how to identify and present key method characteristics of smoothing operators to aid prospective users in making better informed choices of method and parameter values that are suitable for their individual goals and preferences. It is our hope that such an approach would reduce the subjectivity with which edge-preserving filters are currently viewed and evaluated.

\textbf{Identifying the behavioral profile of edge-preserving filters.} The difficulty to assess or characterize edge-aware operators' behavior is a multifaceted problem which has not, as of yet, been resolved. Our proposed methodology is to examine key features of smoothing operators over the entire range of possible smoothing levels and compare between different methods using a common baseline. The importance of identifying a behavioral profile of an operator becomes relevant when evaluating their outputs for selecting a suitable operator for some task or for comparison to other operators. In the literature, such comparisons are often illustrated by either the effect on a 1D signal representing an edge in the input image, image patches or full-sized images. All of these approaches are selective in nature due to their dependency on the input image that is chosen for comparison and their focus on localized spatial effects. In contrast, our proposed approach is both input and locale independent, thus allowing for a more objective and more general evaluation.

\textbf{Primary parameter monotonicity, flexibility and range.} It is intuitive to assume that increase in primary controlling parameter values leads to stronger smoothing of the input image. Indeed, monotonicity in parameter progression is critical for method usability, as a user will often perform a manual semi-binary search in order to achieve the desired level of smoothing. However, our results highlight a complementary behavior: the fact that edge-preserving filters vary in their incremental effect at different parameter values; greater changes in output are often observed at the lower range of parameters. This behavior is counter-intuitive when attempting to fine-tune parameters for a specific smoothing effect and requires users to alter the precision at which parameter values should be specified for different smoothing levels.

\textbf{Characteristic profiles of smoothing operators.} We've found that smoothing algorithms can be characterized in an input-independent manner by the nature of their progressive gradient elimination at increasing parameter values. Such analysis sheds light on the style of smoothing that is achieved by the method (e.g. piecewise constant vs. ``Gaussian-like'') and allows us to group comparable methods by similarity in how they treat gradients in the image. Since it is often the case that the user is not closely familiar with many different kinds of smoothing operators, a grouping of methods similar in output style may be of substantial benefit (e.g., for visual prediction of the smoothing effect).

\textbf{Future work.}
The methodology in this paper would benefit from an extension that includes secondary controlling parameters, which will refine the iterative search results in terms of the choice of equivalently smoothed images. Alternatively, an iterative search procedure may be applied to approximate a combination of features, such as both smoothing and edge-preserving scores. Furthermore, a similar approach can be taken for evaluating application-specific outputs of smoothing algorithms. For example, one may compare images that have undergone texture removal or detail enhancement with different underlying smoothing inputs. Such an approach will demonstrate the direct effect of smoothing operators in a wider context of applications in which they are used.

\bibliographystyle{ieeetr}
\bibliography{EPF-eval-bib.bib}

\clearpage

\end{document}